\documentclass[12pt]{spieman}  

\usepackage{amsmath,amsfonts,amssymb}
\usepackage{graphicx}
\usepackage[colorlinks=true, allcolors=blue]{hyperref}
\usepackage{makecell}
\usepackage[table]{xcolor}
\usepackage{setspace}
\usepackage{tocloft}
\usepackage{lineno}

\title{Ground calibration plan for the Athena/X-IFU microcalorimeter spectrometer}

\author[a,*]{Alexeï Molin}
\author[a,**]{François Pajot}
\author[c]{Marc Audard}
\author[d]{Marco Barbera}
\author[b]{Sophie Beaumont}
\author[b]{Edoardo Cucchetti}
\author[e]{Matteo d’Andrea}
\author[b]{Christophe Daniel}
\author[f]{Roland den Hartog}
\author[g]{Megan E. Eckart}
\author[h]{Philippe Ferrando}
\author[f]{Luciano Gottardi}
\author[i]{Maurice Leutenegger}
\author[e]{Simone Lotti}
\author[e]{Lorenzo Natalucci}
\author[b]{Philippe Peille}
\author[f]{Jelle de Plaa}
\author[a]{Etienne Pointecouteau}
\author[i]{Scott Porter}
\author[j]{Kosuke Sato}
\author[k]{Joern Wilms}
\author[b]{Vincent Albouys}
\author[a]{Didier Barret}
\author[l]{Massimo Cappi}
\author[f]{Jan-Willem den Herder}
\author[e]{Luigi Piro}
\author[f]{Aurora Simionescu}

\affil[a]{IRAP, Université de Toulouse, CNRS, 9 Av. du Colonel Roche, 31400 Toulouse, France}
\affil[b]{CNES, 18 Av. Edouard Belin, 31400 Toulouse, France}
\affil[c]{Département d’Astronomie, Université de Genève, Chemin Pegasi 51, CH-1290 Versoix, Switzerland}
\affil[d]{Università degli Studi di Palermo, Dipartimento di Fisica e Chimica - Emilio Segrè, Via Archirafi 36, 90123 Palermo, Italy and INAF/Osservatorio Astronomico di Palermo G.S.Vaiana, Piazza del Parlamento 1, 90134 Palermo, Italy}
\affil[e]{Istituto di Astrofisica e Planetologia Spaziali, Via Fosso del Cavaliere 100, 00133, Roma, Italy}
\affil[f]{SRON, Netherlands Institute for Space Research, Niels Bohrweg 4, 2333 CA Leiden, The Netherlands}
\affil[g]{Lawrence Livermore National Laboratory, 7000 East Avenue, L-509, Livermore CA 94550, United States}
\affil[h]{Département d’Astrophysique, UMR AIM, Orme des Merisiers, Bat 709, 91191 Gif sur Yvette , France}
\affil[i]{NASA/Goddard Space Flight Center, 8800 Greenbelt Rd., Greenbelt, MD 20771, United States}
\affil[j]{Department of Physics, Saitama University, 255 Shimo-Okubo, Sakura-ku, Saitama, 338-8570, Japan}
\affil[k]{Astronomical Institute of the FAU, Erlangen Centre for Astroparticle Physics, University of Erlangen-Nüremberg, Sternwartstr. 7, 96049 Bamberg, Germany}
\affil[l]{INAF, Osservatorio di Astrofisica e Scienza dello Spazio, via Gobetti 93/3, 40129, Bologna, Italy}

\begin{document} 
\newcommand{\AM}[1]{\textcolor{black}{#1}}

\maketitle

\begin{abstract}
The X-ray Integral Field Unit is the X-ray imaging spectrometer on-board one of ESA's next large missions, Athena. Athena is set to investigate the theme of the Hot and Energetic Universe, with a launch planned in the late-2030s. Based on a high sensitivity Transition Edge Sensor (TES) detector array operated at very low temperature (50~mK), X-IFU will provide spatially resolved high resolution spectroscopy of the X-ray sky in the 0.2-12~keV energy band, with an energy resolution goal of 4~eV up to 7~keV [3~eV design goal]. This paper presents the current calibration plan of the X-IFU. It provides the requirements applicable to the X-IFU calibration, describes the overall calibration strategy, and details the procedure and sources needed for the ground calibration of each parameter or characteristics of the X-IFU.
\end{abstract}


{\noindent \footnotesize\textbf{*}A. Molin,  \linkable{alexei.molin@irap.omp.eu} }
{\noindent \footnotesize\textbf{*}*F. Pajot,  \linkable{francois.pajot@irap.omp.eu} }

\newcommand{\aap}{Astronomy \& Astrophysics}

\newcommand{\apj}{Astrophysical Journal}

\newcommand{\pasj}{Publ. of the Astronomical Society of Japan}

\section{Introduction}
\label{sec:intro}  

\indent Planned for launch in the late-2030s, Athena is ESA's second next large mission, dedicated to the investigation of the Hot and Energetic Universe by answering two main questions, how large scale structures assembled and evolved to the ones we see today, and how black holes grew and shaped the observable universe. 
The Athena X-ray telescope, based on a 12~m focal length Silicon Pore Optics (SPO) mirror, will have two focal plane instruments, the X-ray Integral Field Unit (X-IFU) and the Wide Field Imager (WFI)\cite{2013arXiv1306.2307N, 10.1117/12.2560507}. The Athena satellite will be placed at the Lagrange L1 point in the Earth-Sun system.\\ 
\indent The X-IFU is a microcalorimeter X-ray imaging spectrometer\cite{2023ExA....55..373B}. Its present design\cite{peille_x-ray_2025}, which follows the reformulation of the Athena mission into NewAthena, is based on an hexagonal array of 150\AM{4} Transition Edge Sensors (TES) cooled to 55~mK to obtain high spectral resolution images of the sky. \AM{Such a low temperature is achieved by radiative cooling from the platform down to 50\,K followed by an active cryocooler down to 4\,K and finally routinely cycled to 50\,mK using a multi-staged Adiabatic Demagnetization Refrigerator (ADR). The instrument field of view equivalent diameter is 4 arcminutes.} Transition Edge Sensors rely on the superconducting transition of a material, a combination of Mo and Au in the case of X-IFU \cite{2016SPIE.9905E..2HS, 9362304}, using the steep slope of the resistance of the material as a function of temperature in the transition. The temperature rise caused by the absorption of an X-ray photon on the square absorber laying on the TES  \AM{(317$\mu$m pitch size)} is read as a decrease in current when the TES is operated under constant voltage. This drop in current is amplified by several stages, using Superconducting Quantum Interference Devices (SQUIDs) \cite{2021ITAS...3160356K} at lower temperature stages and a low noise amplifier, part of the Warm Front End Electronics (WFEE) \cite{2020SPIE11444E..3UP} at higher temperature. The final stage of the readout is done with the Digital Readout Electronics (DRE) \cite{2018SPIE10699E..4VR}. The TES properties, such as geometry, materials, electrical and thermal couplings are optimized for the best performances in the soft X-ray energy band, 0.2-12~keV. In order to minimize the thermal load of the system at the lowest temperature stage, the readout is multiplexed, using Time Division Multiplexing (TDM) \cite{doriese_developments_2016, 8671478}. \AM{The current energy resolution of the instrument is expected to be better than 4\,eV below 7\,keV and better than 5\,eV at 10\,keV. The current design target at instrument is 3\,eV at 7\,keV to account for possible performance degradation due to the satellite platform after instrument delivery.} \\
\indent The X-IFU instrument is developed by an instrument consortium composed of more than 50 laboratories or institutions in 12 countries, it is led by IRAP and under CNES technical management. The ground calibration plan, elaborated by the X-IFU Calibration Team (XCaT), takes as input the calibration requirements set at instrument level, deriving from the Athena requirements. It defines the strategy of the measurements to be performed and plans the resources needed, including time, both on ground and in-flight. Among parameters that need to be calibrated, we detail here the strategy that is proposed for five critical quantities : the energy resolution, quantified by the Full Width at Half Maximum (FWHM) of the gaussian core of the Line Spread Function (LSF), the energy scale, relating the pulse height amplitude (PHA) of a pulse to the energy of the incoming photon, and the instrument quantum efficiency (QE), defined as the fraction of incoming photons which end up as detected events. To these quantities, two are added, the background knowledge, which is the expected background level during the observing time in flight, and the timing calibration, which quantifies the error in the timing of detected events.\\
\indent The following calibration plan inherits substantially from the Hitomi/SXS and XRISM/Resolve intruments calibrations \cite{10.1117/1.JATIS.4.2.021406,10.1117/1.JATIS.4.2.021407, 2018PASJ...70...18K, 2018PASJ...70...19M, 2018PASJ...70...20T}. Hitomi was a JAXA mission carrying three instruments on board, among which SXS, a 6x6 pixels microcalorimeter spectrometer, relying on high impedance thermometers \cite{10.1117/12.2055681}. The XRISM mission \cite{2020arXiv200304962X} was developed and successfully launched in 2023 to recover the premature termination of the Hitomi mission. It carries on-board the Resolve instrument, similar to SXS. Since X-IFU also relies on a low temperature microcalorimeter array, an important fraction of its calibration will be based on the knowledge and experience acquired and the hardware developed during the calibrations of SXS and Resolve. \\
\indent We first recall the requirement for the calibration of X-IFU in Sect.~\ref{sec:cal_reqs}. In Sect.~\ref{sec:cal_strat} we outline the calibration strategy, expanding on the energy scale calibration in Sect. \ref{subsec:ES_cal}, the energy resolution in Sect.~\ref{subsec:ER_cal}, the instrument efficiency in Sect.~\ref{subsec:QE_cal}, the background knowledge in Sect.~\ref{subsec:bkg_cal} and the timing calibration in Sect.~\ref{subsec:t_cal}. Finally we provide the specifications and a description of the sources required for the calibration in Sect.~\ref{sec:sources}.

\section{Calibration requirements}
\label{sec:cal_reqs}

In Table~\ref{tab:calib_reqs} we present the requirements for the calibration of X-IFU. The values given refer to the maximum allowed error on the given parameter knowledge. In flight, because of Poisson statistics, different events in the same pixels will not be processed the same way and will receive different “grades” (as many as 6 grading options are considered for now). Throughout this paper, we only address high resolution events, labeled as 'high grade'. \\
\indent The instrument efficiency refers to the probability to detect a photon arriving at the instrument entrance over the entire energy band, which is measured per sub-system. Within the X-IFU requirements, the so-called absolute calibration refers to the knowledge of the average instrument efficiency over the field of view for the specified energy range. The relative calibration refers to the maximum allowed error on the difference in spectral shape across all energies and across the array. 
\begin{table}[ht]

\caption{Summary of the in-flight calibrations requirements for X-IFU. \AM{$\dagger$ : A requirement above 7\,keV has been formulated within the instrument team and is kept as a goal on a best effort basis.}} 

\label{tab:calib_reqs}
\begin{center}    
\resizebox{\textwidth}{!}{
\begin{tabular}{|l|c|c|c|} 
\hline
\rule[-1ex]{0pt}{3.5ex} \cellcolor{gray} \color{white} Element & \cellcolor{gray} \color{white} Knowledge & \cellcolor{gray} \color{white} Applicability& \cellcolor{gray} \color{white} Comment\\
\hline \hline
\rule[-1ex]{0pt}{3.5ex} \cellcolor{lightgray} Energy scale & \cellcolor{lightgray} & \cellcolor{lightgray} & \cellcolor{lightgray} \\
\hline 
\rule[-1ex]{0pt}{3.5ex}  High Grade events  & 0.65~eV (goal 0.5~eV)& 0.2 - $7^\dagger$~keV & \makecell{$1\sigma$ of residuals \\ over 5~ks calibration period}\\
\hline \hline
\rule[-1ex]{0pt}{3.5ex} \cellcolor{lightgray} Energy resolution &\cellcolor{lightgray} & \cellcolor{lightgray} & \cellcolor{lightgray} \\
\hline 
\rule[-1ex]{0pt}{3.5ex} High Grade events & \makecell{6\% of high grade \\ FWHM resolution} & 0.2 - 12~keV  & \makecell{$1\sigma$ of residuals} \\
\hline \hline 
\rule[-1ex]{0pt}{3.5ex} \cellcolor{lightgray} Instrument efficiency & \cellcolor{lightgray} & \cellcolor{lightgray} &\cellcolor{lightgray} \\
\hline 
\rule[-1ex]{0pt}{3.5ex} Absolute cal. & 4\% & 0.2 - 12~keV & \makecell{$1\sigma$ of residuals \\ Same for each pixel  \\ of the array} \\
\cline{2-4}
\rule[-1ex]{0pt}{3.5ex} Relative cal. & 3\% of IE or 0.01 when IE $<$ 0.33 & 0.2 - 12~keV & At pixel level\\
\cline{2-4}
\rule[-1ex]{0pt}{3.5ex}  Edges cal. & 3\% (ground calibration) and 2\% (flight calibration)& 0.2 - 12~keV  & \makecell{At least C, O, N, Al \\ K-edges, \\Au and Bi L-edges, \\ and Au and Bi M-edges}\\
\hline \hline 
\rule[-1ex]{0pt}{3.5ex} \cellcolor{lightgray} Background & \cellcolor{lightgray} & \cellcolor{lightgray} & \cellcolor{lightgray} \\
\hline 
\rule[-1ex]{0pt}{3.5ex}  Background knowledge & \makecell{5\% of the number of non-focused  charged particles }& 100~ks, 9 $\text{arcmin}^2$, $>$1~keV  & -\\
\hline \hline
\rule[-1ex]{0pt}{3.5ex} \cellcolor{lightgray} Timing & \cellcolor{lightgray} & \cellcolor{lightgray}  &\cellcolor{lightgray}  \\
\hline
\rule[-1ex]{0pt}{3.5ex}  Absolute & 5 $\mu$s & 3 $\sigma$, 50~ks  & - \\
\cline{2-4}
\rule[-1ex]{0pt}{3.5ex}  Relative & 10 $\mu$s & 1 $\sigma$, 50~ks & - \\
\hline 
\end{tabular}
}
\end{center}
\end{table}

\section{Calibration strategy}
\label{sec:cal_strat}

\subsection{Overall calibration strategy}
\label{subsec:cal_overall}

\indent The calibration strategy is organized in five parts, each corresponding to the calibration of a given parameter or characteristic of the X-IFU. Corresponding subsections presented below includes a brief description of the procedures as well as the sources, facilities and components needed to achieve a complete calibration of the parameter. \\

The X-IFU calibration strategy combines measurements and analysis:
\begin{itemize}
\item At component level: detector array, filters, readout electronics
\item At sub-system level: during testing within the focal plane assembly 
\item At instrument level, integrated in the Thermal Ground Support Equipment cryostat (TGSE), planned to take place at CNES: this is the main and most complete calibration phase. The full detection chain must be in a configuration as representative as possible of the flight one. A description of the TGSE can be found in section \ref{subsec:tgse}.
\item During Thermal Vacuum/Thermal Balance tests of the the Payload Compartment (PLC) which contains all the instruments (and that will be later integrated with the telescope and the satellite service module):  this is the only opportunity to have a cold instrument using the complete flight cooling chain including passive cooling and all the flight cryocoolers before launch. This phase will take place at the facility selected by the prime manufacturer of the PLC. It will only allow a limited set of measurements and verifications.
\item During flight using internal sources (Modulated X-ray Source (MXS) or $^{55}$Fe radioactive source) or specific X-IFU configuration (filter wheel in closed position).
\item In flight on sky sources.
\end{itemize}
In addition, it will rely on X-ray emission fundamental physics modeling and ground measurements, at high energy resolution. These include simulations or revised and updated standards, as needed for such a high-resolution spectroscopy mission. \\
It should be noted that no end-to-end ground calibration of the X-IFU using the Athena mirror is planned: when needed, the mirror calibration data will be used before flight to compute predicted values that will be checked on real observations of sky sources in-orbit.
It is not planned to bring the X-IFU in front of a synchrotron beam given the complexity of such a test. Instead, we will rely on a set of measurements at sub-system level (e.g. the filters for the instrument quantum efficiency determination).

\indent Table~\ref{tab:calib_stra} shows the strategy that will be followed for the calibration of the various parameters. 

\begin{table}[ht]

\caption{Calibration strategy for X-IFU. A check symbol (\checkmark) indicates when calibration activities are performed. Text entries indicate the component or sub-systems that are considered or other specific characterization or action. The red color highlights the final calibration of a given parameter. \textit{Italics} refers to test or verification activities rather than calibration.} 
\label{tab:calib_stra}
\begin{center} 
\resizebox{\textwidth}{!}{
    \begin{tabular}{|l|c|c|c|c|c|c|c|} 
    \hline
    \rule[-1ex]{0pt}{3.5ex}  ~~~~~~~~~~~~~~~Level & Component & Subsystem & X-IFU in  & X-IFU on   & X-IFU  & X-IFU  & Modeling and  \\
    \rule[-1ex]{0pt}{3.5ex}     &  &  & TGSE  & PLC during& in flight & in-flight & laboratory  \\
    \rule[-1ex]{0pt}{3.5ex}     &  &  &  cryostat & TV/TB tests & \textit{on-board} & \textit{sky} &  physics \\
    \rule[-1ex]{0pt}{3.5ex}   Parameter  & & &  &  & \textit{sources} & \textit{sources} &   \\
    \hline \hline
    \rule[-1ex]{0pt}{3.5ex}  Energy Scale  & \checkmark & \checkmark & \textcolor{red}{\checkmark} &  \textit{spot} & \textit{gain}& cross- & update \\
    \rule[-1ex]{0pt}{3.5ex}    & detector & FPA &  & \textit{check} & \textit{correction}& check & X-ray \\
    \rule[-1ex]{0pt}{3.5ex}    & array & readout &  & & &  & databases \\
    \hline
    \rule[-1ex]{0pt}{3.5ex}  Energy resolution & \checkmark & \checkmark & \checkmark &  \textit{spot} & \textcolor{red}{\checkmark} & cross- & update \\
    \rule[-1ex]{0pt}{3.5ex}   & detector & FPA + GSE & ground & \textit{check} &  & check & X-ray  \\
    \rule[-1ex]{0pt}{3.5ex}    & array & readout & reference & &  &  & databases  \\
    \rule[-1ex]{0pt}{3.5ex}    &  &  &  &  &  &  &   \\
    \rule[-1ex]{0pt}{3.5ex}  Extended LSF & \textcolor{red}{\checkmark} &  & \textit{spot} &  &  &  &   \\
    \rule[-1ex]{0pt}{3.5ex}     & \textcolor{red}{detector} &  & \textit{check} &   &  &  &   \\
    \rule[-1ex]{0pt}{3.5ex}     & \textcolor{red}{array} &  &  &  &  &  &   \\
    \rule[-1ex]{0pt}{3.5ex}     &  &  &  &  &  &  &   \\
    \hline
    \rule[-1ex]{0pt}{3.5ex}   Instrument & \textcolor{red}{\checkmark} &  & \textit{spot} &  & \checkmark & \textcolor{red}{\checkmark}  &   \\
    \rule[-1ex]{0pt}{3.5ex}     efficiency & \textcolor{red}{detector}&  & \textit{check} &  &  & \textcolor{red}{including} &   \\
    \rule[-1ex]{0pt}{3.5ex}      & \textcolor{red}{witness samples}&  & &  &  & \textcolor{red}{telescope} &   \\
    \rule[-1ex]{0pt}{3.5ex}    & \textcolor{red}{\checkmark} &  &  &  &  &  &   \\
    \rule[-1ex]{0pt}{3.5ex}     & \textcolor{red}{filters} &  &  &  &  &  &   \\
    \rule[-1ex]{0pt}{3.5ex}  Contamination  &  &  & pre-launch &  &  & \textcolor{red}{\checkmark} &   \\
    \rule[-1ex]{0pt}{3.5ex}     &  &  & reference &  &  & \textcolor{red}{monitoring} &   \\
    \hline
    \rule[-1ex]{0pt}{3.5ex}  Background & irradiation & \textit{irradiation} & \textit{irradiation} &  & \textcolor{red}{\checkmark} &  & GEANT4 \\
    \rule[-1ex]{0pt}{3.5ex}    & modeling & \textit{check} & \textit{check} &  &  &  & physics \\
    \rule[-1ex]{0pt}{3.5ex}    &  & \textit{FPA} &  &  &  &  & validation \\
    \hline
    \rule[-1ex]{0pt}{3.5ex}  Timing &  & \checkmark & \checkmark & \textcolor{red}{\checkmark} & check & check &   \\
    \rule[-1ex]{0pt}{3.5ex}    &  & readout & readout &  &  & absolute &   \\
    \rule[-1ex]{0pt}{3.5ex}    &  & MXS & MXS &  &  & time &   \\
    \hline  
    \end{tabular}
    }
\end{center}

\end{table}

\subsection{Energy Scale Calibration}
\label{subsec:ES_cal}

\indent The energy scale is the function relating the estimated pulse height of a detected event to the energy of the corresponding photon \cite{10.1117/1.JATIS.4.2.021406, 10.1117/1.JATIS.4.2.021407, 10076800, 10.1117/12.2312170}. It is approximated by a high (8+) order polynomial, and has to be fit using a set of reference points. In other words, constraining the energy scale requires observing a set of known X-ray lines \AM{(10+ for an order 8 polynomial)} covering the entire energy band of the instrument, \AM{ideally evenly spaced in energy}. One implication of this is that an ex-nihilo in-flight calibration of the energy scale is likely to be impossible. For this reason, the plan relies on extensive calibration of the energy scale on the ground  across a large instrument setpoint parameter space for a reliable approximation in-flight. 

\indent The energy scale is sensitive to the operating conditions of the detectors, mainly the thermal bath temperature, the bias voltage, the electronics gain stability, the radiative loads onto the detectors and the surrounding magnetic field \cite{porter2016temporal,2018JLTP..193..931C}. These quantities can be controlled but are likely to drift over long periods of time and to change from one cold cycle to another. This has three consequences :
\begin{itemize}
    \item First, it raises the need for a source capable of illuminating the detector array with several X-ray lines with a sufficient number of counts. 
    A typical X-ray K-shell complex needs between 2,500 and 5,000 counts to be properly sampled at the X-IFU resolution to ensure $\sim$100 meV centroid accuracy, a quantity required to reach the expected constraints needed on the energy scale calibration residuals. For this purpose, we use the Rotating Target Source (RTS) which shines individual samples placed on a rotating wheel and relies on the X-ray fluorescence of the samples to produce lines at various energies \cite{10.1117/1.JATIS.4.2.021406}. The RTS will be used for the energy scale calibration and will allow a complete coverage of the X-IFU energy range on the ground. 
    \item A second consequence is that, as the exact operating conditions encountered during the flight and their potential evolution are unknown, a wide enough range of conditions has to be explored on the ground in order to cover the possible in-flight gain scales. The current strategy is to cover at least 6 possible operating points within the temperature, magnetic field and bias voltage space on the ground to provide 6 gain scales from which the flight gain scale will be derived\cite{10076800, 2018JLTP..193..931C}. The parameter space is only probed for certain parameters, and their values are 5 to 10 times larger than expected to cover for joint drifts in all parameters. \AM{As the operating parameters are correlated in their effect on the gain scale, 6 operating points are expected to be sufficient\cite{cucchetti_advanced_2024, witthoeft_performance_2025}}. The number of operating points is limited by time constraints, as one set point is expected to require roughly 100ks. The goal to reach for the ground calibration curve is a 1$\sigma$ error of 0.15~eV over the desired bandpass. \AM{This value is a rescaling of the 0.1 eV goal before the instrument reformulation and used as a baseline in the energy scale calibration supporting papers\cite{10076800}. This value is deemed achievable given early demonstration activities at NASA.} 
    \item A third consequence is that the ground calibration curves should be taken in an environment more stable than the one expected in flight \AM{to provide accurate reference curves at each setpoint.}
\end{itemize}

\indent The in-flight correction of the energy scale will rely on the previously determined gain scales, and on an on-board Modulated X-ray Source (MXS). The MXS is an X-ray source with a fast response time that produces Brehmsstrahlung of accelerated electrons on a target, together with \AM{characteristic X-ray} emission lines \AM{from the} target (Si, Cr and Cu lines expected). \AM{The Si, Cr, and Cu K$\alpha$ lines will be used for the in-flight correction. }Additional housekeeping data (HK) will provide an estimation of the environment surrounding the detectors, and the quantities affecting the gain scale. A number ($\sim10$) of different housekeeping probes (e.g. temperature sensors) is expected to be used to further refine the correction. Provided the gain scales calibrated on ground cover a sufficiently large parameter space of operating conditions, the in-flight gain scale could be recovered at all times using an interpolation of previous gain scales on the new operating point \cite{2018SPIE10699E..4MC}.\AM{The uncertainties induced by these methods are accounted for in the error budget of the instrument performance.} The short term ($\sim$5~ks) drifts will be monitored by periodically shining the MXS on the detector array (flashes of $\sim$1 s every $\sim$66 s.). Long term drifts, on the other hand, will be verified with the $^{55}$Fe source as well as well chosen sky sources, through ground analysis of systematics and science data.

\indent Energy scales relevant to the flight-like configuration can only be obtained with an instrument close to its final configuration. However, the energy scale(s) will be checked at all possible levels beforehand, beginning at the detector level, albeit with a readout chain different than the one which will be on the final instrument. In the following section, we only describe the calibration procedures that will be run, first, at the Engineering Model (EM) level, with a first flight-like configuration, and then at the Flight Model (FM) level for a final full calibration. It should be noted that the following statements assume an RTS that provides lines known to better than a few tenths of~eV. Then, the ground calibration sequence should be as follows : 
\begin{enumerate}

    \item At the EM level, in the TGSE cryostat, the various energy scale correction strategies will be explored. To achieve that, at least 6 different points in the environment parameters space will be explored. Since this model will not fly, a simplified approach will be led at this stage, in three steps.
    \begin{enumerate}
        \item A first check using a handful of X-ray lines (with one or several MXS specifically developed for ground calibration) to test the energy scale correction for three different physical parameters. This is performed to determine possible departures of the system’s energy scale from the sub-system one.
        \item The proper calibration of the energy scale using six set-points, using the RTS in combination with a fiducial line (MXS or radioactive source).
        \item A third phase, mirroring the first one, where a handful of X-ray lines (MXS) are used to test the energy scale correction for three different physical parameters (two set-points for each). Additional verification of the usability of HK probes as a gain scale correction will be led in parallel.
    \end{enumerate}
    
    \item At the FM level, we will access the energy scale of the instrument that will actually fly.\AM{The FM calibration will be run in the same cryostat as the EM unit, and no major difference in the environment is expected.} Again, three steps will be performed
    \begin{enumerate}
        \item A first check of the curves with the MXS on three setpoints previously measured at subsystem level. 
        \item The proper calibration of the energy scale using six set-points, using the RTS in combination with a fiducial line (MXS or radioactive source).
        \item A flight-like correction, using the MXS, to test the energy scale correction for all physical parameters except linear drift (three set-points for each). In parallel, HK probes shall also be tested to verify the HK sub-system calibration.
    \end{enumerate}
    
    \item At the Payload Compartment (PLC) level, during Thermal Vacuum/Thermal Balance (TV/TB), the instrument will not be accessible with sources such as the RTS. The flight MXS could however be used in the way it will be in flight, allowing a verification of whether the instrument setpoint, in its final configuration, remains in the parameter space that was calibrated in the TGSE. This check will however be done with the Dewar door closed, and the MXS will shine through the beryllium windows in the door.
\end{enumerate}

\indent As mentioned earlier, between 2,500 and 5,000 counts will be needed to ensure the required accuracy over a typical X-ray K-shell at the X-IFU resolution. Considering all the lines in the RTS target with a typical count rate between 0.5 to 1 cts/s/pix, the typical run time for a full energy scale calibration is the duration of a cold cycle (100~ks). Once the energy scale curves have been determined, a lower integration time can be used for spot checks of step 1 and 3 of the strategy at the EM and FM levels. These can be performed using a few lines (typically three) observed with sufficient statistics, i.e. during $\sim$ 5 – 10~ks for typical count rates. With \AM{these} assumptions in hand and accounting for a 20\% margin, a rough estimate of the time required is $\sim$10 cold cycles at the EM level and $\sim$20 at the FM level. At PLC level, one week is expected to be available for instrument verification (beyond PLC level tests). Among these, at least one cold cycle is expected to be dedicated to checking the gain scale at a handful of energies with the MXS and $^{55}$Fe source. This drives the total calibration time at the FM level.

\indent The in-flight calibration will rely on the previously determined gain scales, and the observation of the on-board MXS and radioactive $^{55}$Fe source, as well as the knowledge of the HK data, and possible celestial sources as well (such as Capella for instance). The weighting between use of HK data and MXS data in the final correction technique, as well as their respective order (HK correction applied either before or after the MXS correction) remains to be investigated in later phases of the project. The early post-launch verification will rely on shining the MXS through the (still) closed dewar door, at 6 different set points on at least 3 different X-ray lines (Cr, Cu from the MXS and Mn from the $^{55}$Fe source), for a total of 300ks of observation. If some of the MXS lines were not to have sufficient counts with the Dewar door closed, the MXS checks should wait for the door to be opened. Once the early spot-checks are done and the Dewar door opened, the validity of the previous calibration will be checked using astrophysical sources. In case significant gain scale changes are observed with respect to ground conditions, a full campaign of energy scale calibration should be considered. This would represent a long activity unlikely to provide the same level accuracy as would be possible on the ground.

\indent Finally, during its lifetime, the instrument may evolve and age, and new referential energy scale functions may need to be computed using onboard and sky sources. The use of sky sources is not discussed here but is considered and requires further studies to assess its feasibility. First inputs and return of experience from XRISM demonstrate good confidence in reaching the expected levels of knowledge for the energy scale. All lessons learned shall be included in the calibration plan\AM{, see e.g. Sec. 6 in Eckart et al. 2025\cite{eckart_energy_2025}, suggesting that the on-board processor records the pulses height of all high resolution events with both the high resolution and mid resolution modes. This operating mode is planned for X-IFU}.

\subsection{Energy Resolution and Redistribution Calibration}
\label{subsec:ER_cal}

The line-spread function (LSF) of microcalorimeters is dominated by a Gaussian redistribution core, which we refer to interchangeably as the detector energy resolution or the ‘core LSF’. The LSF of typical microcalorimeters can be adequately described by Gaussian broadening over several orders of magnitude down from the peak. At very low levels, the LSF becomes non-Gaussian due to several energy loss mechanisms primarily related to the photon absorption physics; we refer to energy redistribution due to these effects as the ‘extended LSF.’ The Full Width at Half Maximum (FWHM) of the Gaussian core LSF is defined as the energy resolution of the detectors at the given energy. The extended LSF is mainly dependent o\AM{n} the physical properties of the absorbers, meaning its characterization on the ground should remain valid in different operating conditions and in flight. The calibrated LSF will be used to generate the pre-launch Response Matrix File (RMF).

\indent The LSF ground calibration requires monochromatic lines at multiple X-ray energies across the bandpass. To answer this need, rather than relying on fluorescent lines which can show complex substructures and be poorly characterized, crystal monochromators will be used. In particular, the calibration will rely on channel-cut crystal monochromators (CCCMs) to provide monochromatic lines across most of the waveband\cite{leutenegger_simple_2020} . In these CCCMs, a commercial X-ray tube illuminates a pair of channel-cut crystals that are aligned in a dispersive configuration to select a portion of the $\text{K}_{\alpha_1}$ line of the anode material. The result is a narrow line width of $<0.5$~eV.  CCCMs at Cr~K$\alpha$ and Cu~$\alpha$ were used for Astro-H SXS calibration\cite{10.1117/1.JATIS.4.2.021406}, and an expanded suite of sources were used for XRISM Resolve calibration, providing lines from 4.5~keV up to 11.4~keV \cite{2016SPIE.9905E..3UL, 10.1117/1.JATIS.4.2.021406}. 

\indent Improvements and additions to the suite of monochromators is required for X-IFU calibration. First, the channel-cut crystal pairs used for XRISM calibration (4.5--11.4~keV) will be coupled to large spot-size, water cooled X-ray sources. This change will allow approximately half of the X-IFU detector array to be calibrated simultaneously, as opposed to a single row of detectors ($\sim 1/40$ of the array) with the small spot-size X-ray sources used for XRISM calibration. Second, a suite of monochromators for lower energy lines ($\approx 0.5–4$~keV) are under development. These low-energy crystal monochromators will operate in vacuum. They include CCCMs that use alternative crystal material (not solely silicon crystals, as are used in the $E>4$~keV CCCMs) and monochromators that employ four flat crystals, the latter for the lowest X-ray energies. 

\subsubsection{Core LSF calibration}

\indent The calibration of the core LSF will rely on the observation of a set of X-ray lines across the bandpass, and modeling the results for each event grade. This calibration will also need to include an investigation on the dependence on operating conditions, arrival-time correction, and potential effects of high-count rates. It is estimated that estimating the core LSF for a single set-point of operating conditions is achievable in 100~ks, i.e. one cold cycle, with typical fluorescence lines. \AM{This assumes narrow band CCCMs that cover about 1/40th of the X-IFU array, with 200 counts per line per pixel to obtain a 0.2~eV accuracy on the line FWHM, and 1 count/pix/s. }This calibration gains from the error on the broadening being $1/\sqrt{2N}$ rather than $1/\sqrt{N}$ for the centroid with $N$ the number of counts. \AM{Large spot CCCMs are currently under development and will allow to greatly reduce the time required needed for the core LSF calibration. }The measurements\AM{for the core LSF} should first be performed at the detector level, but must be repeated at the Focal Plane Array (FPA) level then at the instrument level in a configuration as close as possible to its configuration in flight since the core LSF depends on the full detection chain and is sensitive to the operating environment. Checks of the core LSF will be performed last before launch during TV/TB of the FM.

\subsubsection{Extended LSF calibration}

\indent The calibration of the extended LSF can occur at any level of assembly where the monochromators can be accommodated \cite{2019ITAS...2903420E}. The extended LSF requires very long exposures of single X-ray lines. Because it relies mainly on detector physics, it should not depend on the operating environment is likely to be calibrated only once, at the detector level. The extended LSF has three main components, which are as follows :
\begin{itemize}
    \item The electron-loss continuum and escape peaks. The former is a result of the scattering of photo-electrons out of the absorber. It is expected to show a constant flux per unit energy and only requires a normalization calibration. The latter is due to the escape of fluorescence photons out of the absorber material. 
    \item The low energy tail, which appeared for photons of energies below $\sim$2~keV as an exponential tail added to the core LSF in the Hitomi/SXS detector array. The behavior of this tail with incident energy suggests that it results from long-lived surface excitation of the absorber material in the case of Hitomi/SXS. The exact mechanism may be different in the case of X-IFU pixels, and may likely be observed above 2~keV, and as such will need to be characterized.
    \item The Silicon K fluorescence, which comes for photons that were absorbed in the underlying Silicon substrate, either because they fell between two pixels or went through one without interaction, occasionally leading to back fluorescence onto the main detector absorbers. The strength of this feature as a function of the incident flux will need to be calibrated.
\end{itemize}

\indent In flight, the LSF will be firstly checked after the launch, and secondly monitored for potential evolution during normal operations. Based on the ground calibration the functional form of the energy resolution versus X-ray energy and its dependence on the system noise will be understood, allowing a re-generation of the core LSF calibration curves if needed, using measurements of the baseline resolution. The monitoring of the core LSF for potential drifts will be made with the MXS and potential astronomical sources. The extended LSF should not change on-orbit, however observations of astronomical sources could be made to validate the models.

\subsection{Instrument Efficiency Calibration}
\label{subsec:QE_cal}

\indent The Instrument Efficiency (IE) is defined as the probability of a photon focused by the mirror toward the X-IFU focal plane location to be detected by the X-IFU microcalorimeter array, such that the total Athena effective area is a combination of the mirror effective area and the X-IFU instrument efficiency. The instrument efficiency is the product of the transmission of the optical blocking filters, the contamination on the blocking filters, and the detector Quantum Efficiency (QE). In practice, it also includes dead time and yield (the fraction of live pixels). 

\indent The X-IFU will have a Filter Wheel (FW) outside of the Dewar, carrying additional filters, mainly for observing the brightest sources. The FW carries one optical filter, designed for blocking the optical flux and one Be filter, designed for blocking the X-ray flux. \AM{These} filters are different from the Thermal Filters (THF), designed to block the infrared radiation, and which are mounted in the Dewar. Each of \AM{these} filters will need to be calibrated. During a part of the ground observations and the first sky observations, the Dewar beryllium door will be shut. The Dewar door transmission will then need to be calibrated as well.

\indent The individual components in the optical path will each be calibrated separately on the ground and these measurements used to calculate the pre-launch total instrument efficency. Calibration of the end-to-end effective area will be performed in-orbit using astronomical sources. 

\subsubsection{Thermal filters calibration}

\indent The X-IFU aperture assembly includes a stack of a four THFs  anchored to the nested temperature stages. Each filter is composed of a thin aluminized polyimide film affixed with epoxy to a $\sim$98\% open-area support mesh. The THFs are designed to transmit X-rays while blocking optical and infrared radiation. The support meshes serve three main purposes: to mechanically support the films, to aid in thermal conduction from the filter frame, which has heaters and a thermometer, throughout the filter film, and to insure EMI shielding for the detectors.\\
\indent Each filter will be calibrated using synchrotron measurements, as well as microscopy observations and X-ray mapping. In addition to the calibrations of the flight filters, witness samples will be purchased, which will be produced in the same batch as the flight filters, as well as Aluminium/polyimide filters with and without a thin film. These will serve as a way to reduce the risks on the flight filters, and the time needed in using those, as well as providing high quality data on the optical constants of the materials involved. 

\indent The calibration strategy is based on the experience in calibrating Hitomi/SXS and XRISM/resolve \cite{10.1117/1.JATIS.4.2.021406,10.1117/1.JATIS.4.2.021407, 2018PASJ...70...18K, 2018PASJ...70...19M, 2018PASJ...70...20T}, and is divided in three (plus one optional) steps :
\begin{itemize}
    \item The calibration of standards, which refers to the accurate knowledge of the optical constants of aluminium and polyimide as a function of energy, as well as all the edges (Al and O K edges in Aluminum, and C,N and O in Polyimide). The possible temperature dependence shall be investigated as well. The knowledge of the edges should be at least a factor of 2-3 better than the X-IFU resolution.
    \item The calibration of the X-IFU witness samples, intended to provide measurements of film component thicknesses without requiring handling of the flight filters. Because the fine-structure measurements will be performed on the calibration standards, measurements with energy steps in the range 5-10~eV over the full energy range, and $\sim$0.8~eV in a narrow energy range (few tens of~eV) around the edges should be sufficient to properly model the filter transmission and derive the areal densities of the atomic elements in the compound with sufficient accuracy.
    \item The calibration of the flight filters, intended to validate witness measurements, test for uniformity, measure fine mesh filling fraction and epoxy filling factor. Spatial scans will be performed to confirm the required filter transmission  uniformity, or to quantify any non-uniformity, with separate measurements of the mesh bars and the deposited film. 
    \item The optical measurements of filter meshes, before film is attached, which is optional but likely to be led, using optical digital microscopy mapping and SEM (Scanning Electron Microscope) measurements to measure the size and shape of the bars and verify the plating quality.
\end{itemize}

\indent In flight, contamination can happen on the filters \cite{2015SPIE.9601E..07O}. Contamination refers to any and all matter that deposits on the filter, most possibly ice that form\AM{s} because of the condensation of residual vapors within or around the Dewar assembly. The X-IFU aperture assembly will be designed to avoid contamination on the inner blocking filters, and the filters heated to avoid condensation when possible. Despite these measures, X-IFU will require on-board monitoring of the contamination, by observing astrophysical sources. Such sources are not confirmed yet, but are likely to be similar to sources used for the SXS and later XRISM calibrations. Among them, we can cite Supernova Remnant 1E0102-72 and the Isolated Neutron Star RXJ1856-3754, which have proven to be stable at better than 1\% level \cite{2012SPIE.8443E..12P,2012A&A...541A..66S}.

\subsubsection{Detector Quantum Efficiency calibration}

\begin{figure} [ht]
\begin{center}

   \includegraphics[width=0.5\textwidth]{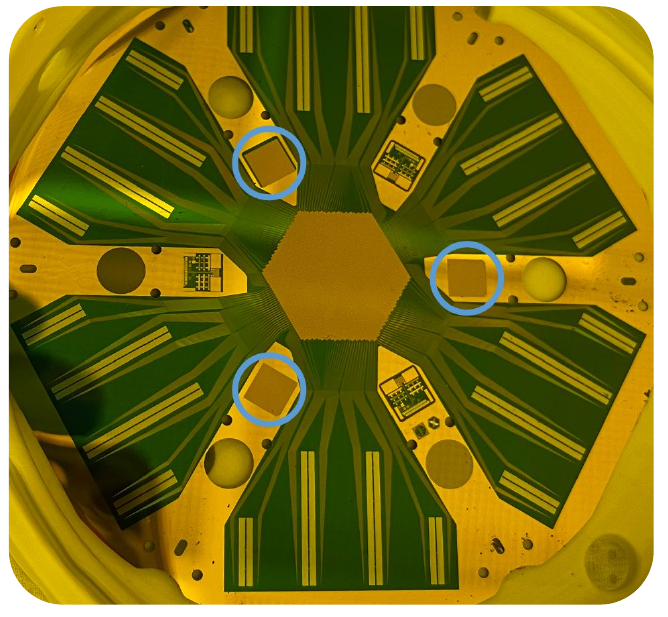}

\end{center}
\caption[example] {\label{fig:TES_witness_sample}Photograph of a prototype X-IFU detector array with absorber witness samples circled in blue. QE measurements of these samples were made at the IDEAS beamline at the Canadian Light Source in 2019 (Eckart, et al., 202\AM{6}, in prep). }
   \end{figure}

\indent The total X-IFU detector Quantum Efficiency (QE) is a combination of the absorber QE and filling fraction of the array \cite{10.1117/1.JATIS.4.2.021406}. The dead pixel fraction is not included in the QE but is expanded upon at the end of this section. The absorber QE refers to the X-ray stopping power of each pixel, which is determined by the X-ray absorber materials and thicknesses. The thickness and composition of the absorbers have been optimized to maximize this quantity over the desired energy range while maintaining appropriate heat capacity for energy resolution. Assuming that these parameters are uniform enough on the array, a single QE will be affected to the detector matrix; however, this will need to be verified and it is possible that the QE will vary slightly over the array and this will need to be reflected in the calibration database. 
The QE of an absorber is computed as a function of the X-ray incident energy E using the following equation :
\begin{equation}
    QE_{\rm{absorber}} = 1 - e^{- \mu_{\text{Au}}(E) \Sigma_{\text{Au}} - \mu_{\text{Bi}}(E) \Sigma_{\text{Bi}} }
\end{equation}
with $\mu(E)$ the mass absorption coefficient and $\Sigma$ the areal density. The mass absorption coefficient\AM{s} are known to a good enough level\cite{1993ADNDT..54..181H} so that the uncertainty is dominated by the measurement of the areal density. In order to calibrate this quantity, during the manufacturing of the detector chip, large absorber witness samples will be fabricated near the array. In Fig. \ref{fig:TES_witness_sample}, we show photographs of a prototype array and highlight the witness samples. The QE calibration will proceed in a three step approach:
\begin{itemize}
    \item During the fabrication of the array, measurements of the mass before and after the deposition of each layer will provide an estimate of the absorber areal density.
    \item By measuring the sample area using high-power microscopy and weighing the witness samples, the areal density can be estimated at three locations on the chip.
    \item Calibrating the X-ray transmission of witness samples at a synchrotron facility will provide a measurement of the QE. Comparisons of the QE at the three witness samples will provide a measure of the non-uniformity of the QE across the detector array. 
\end{itemize}
The expected overall precision over this quantity is a knowledge better than 3\% of the QE over the X-IFU energy bandpass.

\indent The filling fraction is the geometrical proportion of the array area that is covered by absorbers. Because there is a gap between each pixel and side walls which make the pixels not perfectly rectangular, it is not unity. In the current configuration, the pixel pitch is 317$\mu$m with an absorber width of 311.5$\mu$m, giving a pixel filling fraction of the order of 0.96. This quantity will be measured with photographs using high powered microscopy after fabrication and before delivery for further integration in the X-IFU subsystems. It is not expected to change on orbit. The knowledge of the pixel filling fraction is expected to be better than 1\%.

\indent The dead pixel fraction is the fraction of inoperable pixels, and can be easily provided before flight and updated later-on. How to account for pixels that are intermittently bad is not decided yet.

\subsubsection{Filter Wheel Optical Blocking Filters Calibration}

\indent The FW will accommodate one (thin) OBF designed for visible bright sources, one high count rate (thick, Beryllium-like) for bright X-ray sources. The X-ray transmission of each filter and its support structure must be calibrated on the ground, which will include a combination of synchrotron measurements and ray-tracing calculations to correctly model the vignetting due to support structures. \\
\indent The thick filter transmission will be calibrated using synchrotron measurements on a few (approx. 5) locations on the filter, over the entire energy bandpass, with a high energy resolution. The homogeneity of the filter will be assessed with a 50x50 grid of measurements over the piece at a few energies. \\
\indent The thin filter will be made of a aluminium mesh with a polyimide film and are thus likely to be calibrated in the same time and manner as the THFs. \\
\indent The Dewar door will remain closed during all the first operations of X-IFU in orbit, and needs to be calibrated as well. The door includes a Be window, which will follow the same calibration procedure as the thick OBF. The possible metallic support structure of the door will be photographed as well, to check against the plans, and ensure that the design leads to little to no obstruction of the optical path, using ray-tracing.

\subsection{Background knowledge}
\label{subsec:bkg_cal}

\indent The background is defined as all events that produce events as X-ray photons would do, but that do not come from the observed source. It can be sorted into three components :
\begin{itemize}
    \item the Galactic Cosmic Rays interacting with the satellite, with high enough energy (E $>$ 150 MeV) to reach the detectors directly or via secondary showers.
    \item the low energy (E $<$ 100~keV) charged particles (often referred as soft protons) collected by the mirrors. Most of these particles are deflected away from the detector by the charged particles magnetic diverter \cite{8309389}.
    \item in-band X-ray photons coming from unresolved astrophysical sources and diffuse foreground galactic emission.
\end{itemize}
Because these events are not discernible from the ones from the astrophysical sources, this background needs to be quantified precisely. It should be noted that background calibration will not be an instrument only activity, as there are plans to potentially use WFI to scale the background levels to cover for its variability.

\indent Ground activities for the background verification will be firstly focused on modeling and simulations of the expected background on board X-IFU, relying on the knowledge of the expected GCR flux/GCR-induced background, the detailed assembly of the spacecraft and Monte Carlo simulations \cite{2021ApJ...909..111L}. Background events are first vetoed by detection of auto-coincidences on the main array and then by the CryoAC, an anti-coincidence detector placed beneath the X-IFU detector array made of Si-suspended absorbers sensed by IrAu TESes\cite{2023ExA....55..373B} This vetoing process will be checked during the ground calibration. This requires the validation of the CryoAC operation: threshold, energy band and relative timing with respect to the microcalorimeter array and dead time. The calibration of its energy scale will use $^{55}$Fe (6~keV, close to the CryoAC low energy threshold) and $^{241}$Am (60~keV, closer to the typical MIP energy deposition ~150~keV) sources on the ground, together with an on-chip heater to cover the complete energy range. These pre-flight activities will ensure (1) that the requirement on the level of the background undetected by the vetoing process will be met in-flight, and (2) that all effects of high energy particles on the X-ray astrophysical signal are well understood. The full background calibration will only occur in flight, relying on data from the X-IFU microcalorimeter array and the CryoAC, as well as the WFI.


\indent The Non X-ray Background (NXB) is all events that would be detected by X-IFU with the filter wheel in its closed position, hence not caused by X-rays. The NXB is expected to produce a broken powerlaw across the 0.2-12~keV energy range, with a relatively flat part over the 2-12~keV energy range. The current specification foresees an absolute knowledge of the level of the NXB within 5\% for a 100~ks observation over a solid angle of 9 $\text{arcmin}^2$. Three options exist to calibrate the NXB \cite{2018SPIE10699E..4NC}:
\begin{itemize}
    \item Observation with X-IFU only with the filter wheel in closed position, requiring approx. 5~ks to reach desired knowledge
    \item Observation with the high energy band of X-IFU (10 to 15~keV), requiring approx 15~ks, but which do not give access to short time variability of the NXB
    \item Using the CryoAC as a particle flux monitor to infer the level of NXB seen by X-IFU, which remains to be a validated method
\end{itemize}
 Because of temporal variations of the satellite’s environment, the level of NXB may vary significantly between observation phases. Monitoring the NXB should therefore be a regular operation during the non-observing time of the instrument. At the Athena level, it will be possible to add WFI as a monitor of the temporal variations of the NXB when X-IFU will be in use. However, the level of NXB observed by WFI is not expected to be the same for X-IFU, hence only assuming a proportionality between the two.\\
\indent The monitoring of the soft proton background is not within the X-IFU calibration plan as it is driven by the charged particles diverter performance. In other words, given expected diverter performance, the contribution from soft protons is expected to be negligible. If the diverter is less efficient than expected, or there are other unexpected/unaccounted factors, based on the XMM experience there will be an additional rapidly variable component of the particle background.\\
\indent The astrophysical background depends on the position on the sky. When needed it will be estimated in flight from areas of the detector not covered by the source, from off-source pointings whether dedicated or from nearby already astronomical sources.

\subsection{Timing calibration}
\label{subsec:t_cal}

\indent The purpose of the X-IFU timing calibration will be twofold: 
\begin{itemize}
    \item Determine at the instrument level the proper correction for the arrival time of the main detector events so that the overall X-IFU absolute timing accuracy, including Athena contributions, does not exceed 50 $\mu$s.
    \item Conduct a similar effort for the CryoAC detector so that it can issue veto windows towards the main detector when background events are detected. 
\end{itemize}

\indent On the main detector, the timing of each event will be provided by the Event Processor (EP \cite{cobo_athena_2018}). Because of the different filters applied to the different grades of events, the non-linearity of the TES responses, and the different operating conditions of each TES, the timing correction coefficient should depend on the pixel, grade and energy of the event. The calibration sequence will be first made on ground, by shining an MXS on the array with short pulses at a high frequency and using the timing of these pulses to calibrate the timing with the EP. \AM{A dedicated operating mode of the EP will be able to process all pulse grades simultaneously, which will allow to derive the appropriate timing coefficients for all grades. The MXS flashes will be about 50$\mu$s, with a 50ms afterglow. Simulations taking these figures into account show that the short rise time (below 1$\mu$s) of the MXS flashes still allows to reach requirements for the timing ground calibration.} In flight, the final calibration will be obtained with the observation of millisecond pulsars \cite{2017ApJ...845..159G} and comparison with other radio or X-ray observations. 

\indent On the CryoAC, the timing correction can be dependent on the pixel and energy as well. On the ground, its timing capabilities will be calibrated first with natural cosmic rays at ground level (muons), or with the MXS and later maybe with a radioactive source.

\subsubsection{Main detector timing calibration sequence}

\indent Because the final version of the pulse templates used in the EP will be acquired in a close to final configuration of the instrument in order to maximize its performance, i.e. at least when X-IFU is integrated in CNES TGSE level, reliable timing correction data will only be accessible at this stage. In the following description, we present the timing calibration at the different levels, with first tests at the EM level and the real and final calibration at PLC level:
\begin{itemize}
    \item At EM level, given the MXS pulse time, shining 2 counts/s/pixel and the X-IFU expected timing resolution of 10$\mu$s, it is estimated that 10~ks should be sufficient to obtain the proper timing correction coefficients. This stage will however only be a check of the method for the final calibration at PLC level.
    \item At FM level the measurement shall be repeated in order to prove the capability of calibrating at PLC level
    \item At PLC level, since the final timing performance can only be achieved with the proper hardware, i.e. representative warm harnesses, including routing, the final set of time correction coefficients shall be performed during a dedicated cold cycle during TV/TB.
    \item In flight, there are two millisecond pulsars which are promising candidates for timing : PSR J0218+4232 and PSR B1937+21. Both emit very sharp pulses with a large modulation fraction over a large energy band, with a pulsed emission detected even below 0.5~keV for PSR J0218+4232.
\end{itemize}

\subsubsection{CryoAC timing calibration sequence}

\indent For the same reason as for the main array, the final cryoAC timing calibration will be performed at PLC level. The timing calibration at the different levels is described in the following sequence :
\begin{itemize}
    \item At EM/FM level, in the CNES TGSE, the cryoAC relative timing calibration to the main detector can be achieved by using the natural muon flux at ground level. With 0.5 cts/cm2/min, and a timing resolution of 10$\mu$s, a 100ks exposure time would achieve the required timing need. A dedicated $^{241}$Am source will be used in addition for this task.
    \item At PLC level, the previous calibration will once again be performed on a flight-like configuration, using sufficient dark and MXS only exposure time.
    \item In flight, the cryoAC veto timing will be checked during dark observations already required for the background calibration, or with MXS on during closed observations, which should provide enough exposure time to correct any change from the ground.
\end{itemize}

\section{Calibration hardware}
\label{sec:sources}

\indent This section outlines the needs in terms of hardware for the various calibration campaigns

\subsection{The TGSE Dewar} \label{subsec:tgse}

\begin{figure} [ht]
    \begin{center}
    \includegraphics[width=0.85\textwidth]{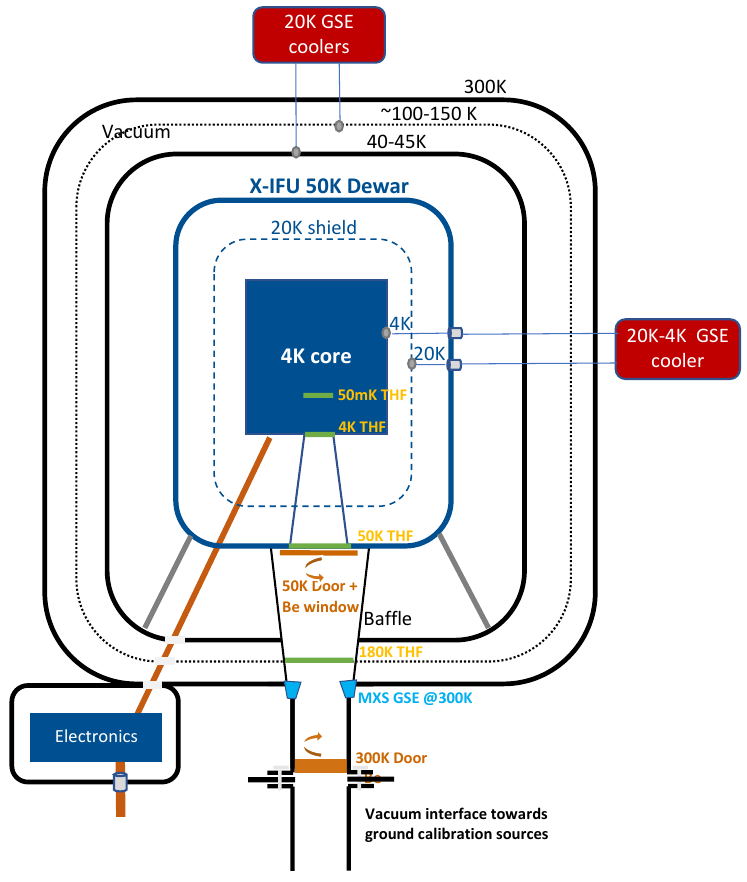}
    \end{center}
    \caption[example] {\label{fig:CGSE} Schematic view of the Testing Ground Support Equipment (TGSE) cryostat.  The Ground Support Equipment (GSE) coolers refer to coolers specifically developed for the TGSE. }
\end{figure}

\indent The main calibration phase of X-IFU will take place in a dedicated cryostat, the TGSE. This cryostat is designed to receive the X-IFU instrument integrated in its own 50 K flight Dewar, and will deliver the 20~K and 4~K thermal interfaces that will be ultimately provided by the X-IFU 4~K flight cooler. The TGSE will provide cold harnesses between the X-IFU 4 K core (detectors, ADR refrigerator and all electronics up to 4 K) and room temperature electronics, similar to the X-IFU flight harnesses, so as to be as representative as possible of the flight configuration.
An optical interface under vacuum will allow to connect X-ray ground calibration sources covering the complete 0.2 – 12~keV energy range to the X-IFU Dewar entrance (Fig. \ref{fig:CGSE}).

\subsection{Portable Channel Cut Crystal Monochromators}

\indent Channel Cut Crystal Monochromators (CCCMs) rely on the Bragg diffraction of X-rays on a crystal. The condition on the diffraction angle and incoming energy to produce constructive interference makes for highly peaked diffraction patterns as a function of angle and high selection of the incoming energy. The achieved monochromaticity is of the order of a few 100s of meV \cite{2020RScI...91h3110L}\\
\indent CCCMs fall into two main categories: low-energy monochromators ($<$3~keV) and high-energy CCCMs to cover energies from 4.5 to 11.4~keV. For the former, the existing version developed for the calibration of XRISM will require a more compact version to accommodate for the down-looking orientation of the planned X-IFU test platforms. For the latter, CCCMs using silicon crystals can generate lines between 4.5~keV and 11.4~keV, depending on the material used for the anode and the associated $\rm{K}_{\alpha_1}$ line. The output beam of each CCCM provides a collimated image of the anode spot in the dispersion direction, and is unfocused in the cross-dispersion direction, so the CCCM beam appears as a line at the detector array, if the anode spot is small. To cover the entire (or at least an acceptable fraction of\AM{the}) X-IFU array however, new high-power water-cooled X-ray tubes are under development, as the idea is to produce a large spot size at the anode, which generates more heat. 

\subsection{Rotating Target Source}

\begin{figure} [ht]
\begin{center}

   \includegraphics[width=\textwidth]{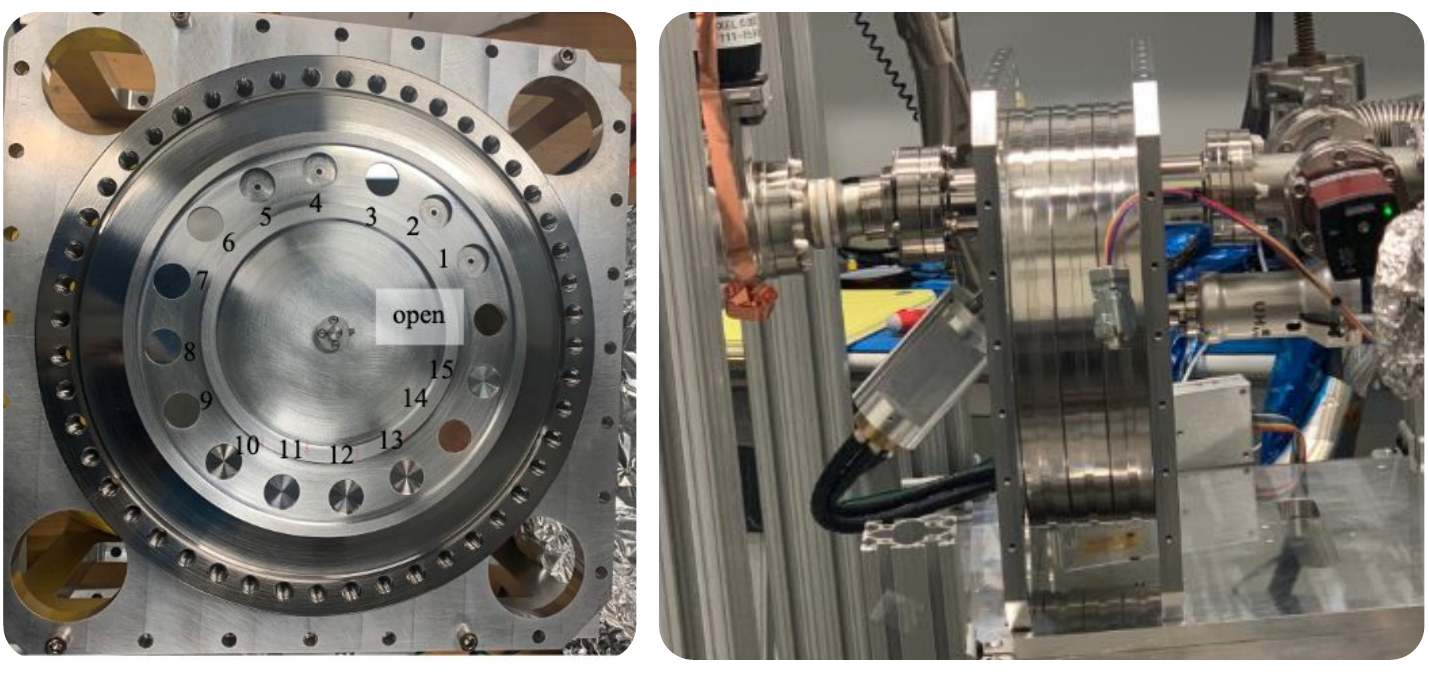}

\end{center}
\caption[example] {\label{fig:RTS} (Left) Photograph of the sample wheel inside the RTS developed at the NASA GSFC, with the numbers of each sample and the open position annotated. (Right) Photograph of this same RTS in operation. The X-ray tube can be seen fixed at an angle with respect to the wheel. The X-rays emitted from the fluorescence are seen from the tube orthogonal to the wheel. }
   \end{figure}

\begin{figure} [ht]
\begin{center}

   \includegraphics[width=10cm]{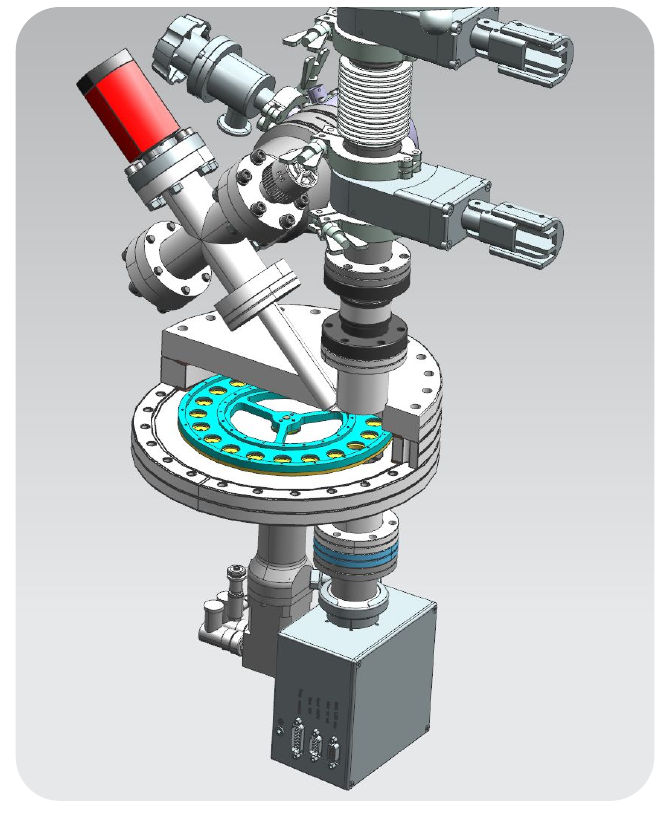}

\end{center}
\caption[example] {\label{fig:RTS_scheme} Section view of the RTS developed for the X-IFU calibration. The view has been cut to allow a view of the wheel, in blue, and the samples carried in yellow. The X-ray source lies in the cross above the angled tube, illuminating the source at a 45 degree angle. Below the RTS is the stepping motor, and the box aligned on the optical axis is the MXS. Above the RTS, on the optical axis, there are isolating sections, valves, and a bellow.}
   \end{figure}

\indent The RTS consists of an X-ray tube illuminating a spot on the radius of a wheel of different material samples. By rotating the wheel, different samples are exposed to the X-rays and shine through fluorescence. In Fig \ref{fig:RTS}, we show photographs of the RTS developed at the NASA GSFC, in Fig. \ref{fig:RTS_scheme} we show a side cut of the planned RTS installation at IRAP. This allows a continuous exposure to lines at very different energies between $\sim$500~eV and $>$15~keV. This range is strongly dependent on the elements picked and the X-ray tube, which needs to provide enough X-ray photons above the desired fluorescent transitions, depending on the anode material and maximum voltage of the source. The RTS is the ideal source for energy scale calibration as it can be set to provide reasonable flux on (quasi) monochromatic lines over the entire X-IFU energy range. \\
\indent There is however one limitation to the RTS, which is the limited knowledge available of the atomic lines, especially at low energies (below $\sim5$~keV). The calibration on these lines can only be done if they are known to an accuracy much better than the instrument energy scale calibration requirement. This is not the case at the present day. It is however possible to cross-calibrate the RTS on a better known source, and this is the planned action to provide a well known RTS for the calibration. Such a source will be an Electron Beam Ion Trap (EBIT). The EBIT is described in Sec.~\ref{subsec:ebit}. 
Operating an array of TES detectors in front of an EBIT simultaneously with an RTS will provide an excellent knowledge of the low-energy lines of an RTS. 

\subsection{Modulated X-ray source}

\indent An MXS relies on a UV diode shining on a photocathode that is sensitive to UV light \cite{2018SPIE10699E..65D, 2018JATIS...4a1204D}. With a proper set of anodes and cathodes, the emitted electrons can be accelerated and demultiplied in number towards a target. If the electron\AM{s} have a high enough energy, their interaction with the material target causes fluorescence lines, as well as Brehmsstralung radiation, in the X-ray range of interest. The MXS does not rely on heating up an anode, and thus benefits from a much lower response time, allowing fast response of X-ray emission. 

\subsection{Electron Beam Ion Trap}
\label{subsec:ebit}
\indent Although not directly needed for the calibration of X-IFU, the EBIT proves to be the only viable option for calibrating the RTS. It should be noted that in principle, it could be possible to shine directly an EBIT on the X-IFU array\AM{, however the expected flux of an EBIT would require prohibitive exposure times. Moreover, the }size and complexity of an EBIT would make such a configuration prohibitively complex. The effort put into calibrating an RTS is paid-up for by the relative simplicity and compactness of the RTS, which will show as a net gain of time and effort when dealing with the calibration of X-IFU in the TGSE. \\
\indent The working principle of an EBIT relies on the ionization of one or several gases in a trap, by an electron beam. The electron beam is generated by an electron gun, and the electrons are accelerated towards the trapping region. This region is a needle a few tens if $\mu$m in radius and a few centimeters in length. It is surrounded by drift tubes, which are biased in voltage to provide a potential that serves two purposes, accelerating the electrons, and trapping the ions along the axial direction. This assembly is surrounded with superconducting Helmholtz coils providing an intense (few Teslas) magnetic field, which focuses the electron beam, and confines the ions radially when the electron beam is off. \\
\indent By tuning the different parameters, such as the electron current and arrival energies, the gas content,  etc, one can access a wide range of electronic transitions within the ionized gas. Mainly, one can reach states where the atoms are stripped from all but one or two of their electrons. Such states are called hydrogen or helium-like, \AM{they are much more easily modeled in theory and have been validated to greater degrees of precision than lower charge states.} This allows a very good knowledge of these transitions in the X-ray band, with a knowledge of the lines of the order of the milli-eV.

\subsection{Synchrotron light sources}

\indent When a relativistic particle is accelerated perpendicularly to its trajectory, it emits synchrotron radiation. This phenomenon is used in synchrotron facilities with relativistic electrons accelerated in a ring as a means of produced well characterized X-ray light. Access to synchrotron light sources for X-IFU calibration is already planned and performed for filters and detector absorber samples measurements. These include so far Bessy (Berlin, Germany), Elettra (Trieste, Italy) and Soleil (Saclay, France), and will also include facilities such as the Stanford Synchrotron Radiation Lightsource (SSRL) (Stanford, United States).

\section{Conclusion}

\indent In this paper, we have reviewed the requirements and the calibration procedures for each of the main X-IFU parameters.
Because of the complexity of operating X-IFU in its final configuration at the PLC level, which will be done under the responsibility of the \AM{satellite manufacturer}, the current calibration plan will allow to meet the calibration requirements with an optimal split between activities carried under the consortium responsibility and the \AM{satellite manufacturer}, within a reasonable amount of time including safety margins.\AM{The current total time required for calibration at the FM level is estimated to be of the order of 3 months and is fully compatible with the instrument schedule.} The performance of X-IFU is tightly linked to the quality with which it will be calibrated, which outlines the importance of a detailed and systematic calibration plan. The plan presented here meets the requirements of the X-IFU, but might be adapted whenever needed as the instrument is being built.

\section*{Disclosures}

\indent The authors declare that there are no financial interests, commercial affiliations, or other potential conflicts of interest that could have influenced the objectivity of this research or the writing of this paper.

\section*{Code and data availability}

\indent Data sharing is not applicable to this article, as no new data were created or analyzed. The code that was used for simulations is kept internal to the X-IFU consortium but can be requested from the author at alexei.molin@utoulouse.fr

\acknowledgments 
 
We would like express our gratitude to the lead funding agencies supporting X-IFU for their support.
The Italian contribution to XIFU is supported by ASI (Italian Space Agency) under Contract 2019-27-HH.0, 2019-27-HH.1-2021, 2019-27-HH.2-2023, and 2019-27-HH.3-2024. Part of this work was performed under the auspices of the U.S. Department of Energy by Lawrence Livermore National Laboratory under Contract DE-AC52-07NA27344. 

\bibliography{biblio} 
\bibliographystyle{spiebib} 
\end{document}